\documentclass[10pt,journal,epsfig]{IEEEtran}

\usepackage[dvips]{graphicx}
\usepackage{graphicx}
\usepackage{amssymb}
\usepackage{cite}
\usepackage{subfigure}
\usepackage{amsmath}
\usepackage{algorithm}
\usepackage{algorithmic}
\usepackage{multirow}
\usepackage{xcolor}

\begin{document}

\title{Compressed Channel Estimation for Intelligent Reflecting Surface-Assisted Millimeter Wave
Systems}

\author{Peilan Wang, Jun Fang, Huiping
Duan, and Hongbin Li, ~\IEEEmembership{Fellow,~IEEE}
\thanks{Peilan Wang, and Jun Fang are with the National Key Laboratory
of Science and Technology on Communications, University of
Electronic Science and Technology of China, Chengdu 611731, China,
Email: peilan\_wangle@std.uestc.edu.cn, JunFang@uestc.edu.cn}
\thanks{Huiping Duan is with the School of Information and Communications Engineering,
University of Electronic Science and Technology of China, Chengdu
611731, China, Email: huipingduan@uestc.edu.cn}
\thanks{Hongbin Li is
with the Department of Electrical and Computer Engineering,
Stevens Institute of Technology, Hoboken, NJ 07030, USA, E-mail:
Hongbin.Li@stevens.edu}
\thanks{This work was supported in part by the National Science
Foundation of China under Grants 61829103 and 61871421.}}

\maketitle

\begin{abstract}
In this paper, we consider channel estimation for intelligent
reflecting surface (IRS)-assisted millimeter wave (mmWave)
systems, where an IRS is deployed to assist the data transmission
from the base station (BS) to a user. It is shown that for the
purpose of joint active and passive beamforming, the knowledge of
a large-size cascade channel matrix needs to be acquired. To
reduce the training overhead, the inherent sparsity in mmWave
channels is exploited. By utilizing properties of Katri-Rao and
Kronecker products, we find a sparse representation of the cascade
channel and convert cascade channel estimation into a sparse
signal recovery problem. Simulation results show that our proposed
method can provide an accurate channel estimate and achieve a
substantial training overhead reduction.
\end{abstract}

\begin{keywords}
Intelligent reflecting surface, millimeter wave communications,
channel estimation.
\end{keywords}

\section{Introduction}
Intelligent reflecting surface (IRS) comprising a large number of
passive reflecting elements is emerging as a promising technology
for realizing a smart and programmable wireless propagation
environment via software-controlled reflection
\cite{HuangZappone19,DiRenzo19,WuZhang19b,BasarDi-Renzo19}. With a
smart controller, each element can independently reflect the
incident signal with a reconfigurable amplitude and phase shift.
By properly adjusting the phase shifts of the passive elements,
the reflected signals can add coherently at the desired receiver
to improve the signal power. Recently, IRS was introduced to
establish robust mmWave connections when the line-of-sight (LOS)
link is blocked by obstructions \cite{TanSun18,WangFang19}. To
reach the full potential of IRSs, accurate channel state
information (CSI) is required for joint active and passive
beamforming. There are already some works on channel estimation
for IRS-aided wireless systems, e.g.
\cite{TahaAlrabeiah19,MishraJohansson19,JensenDe19,HeYuan19,ZhengZhang19}.
In \cite{TahaAlrabeiah19}, to facilitate channel estimation,
active elements were used at the IRS. These active elements can
operate in a receive mode so that they can receive incident
signals to help estimate the BS-IRS channel and the IRS-user
channel. IRSs with active elements, however, need wiring or
battery power, which may not be feasible for many applications.
For IRSs with all passive elements, least square (LS) estimation
methods \cite{MishraJohansson19,JensenDe19} were proposed to
estimate uplink cascade channels. The problem lies in that the
cascade channel usually has a large size. These methods which do
not exploit the sparse structure inherent in wireless channels may
incur a considerable amount of training overhead. In
\cite{HeYuan19}, a sparse matrix factorization-based channel
estimation method was developed by exploiting the low-rank
structure of the BS-IRS and IRS-user channels. The proposed method
requires to switch off some passive elements at each time.
Implementing the ON/OFF switching, however, is costly as this
requires separate amplitude control of each IRS element
\cite{ZhengZhang19}.

In this paper, we consider the problem of channel estimation for
IRS-assisted mmWave systems. To reduce the training overhead,
sparsity inherent in mmWave channels is exploited. By utilizing
properties of the Khatri-Rao and Kronecker products, we find a
sparse representation of the concatenated BS-IRS-user (cascade)
channel. Channel estimation can then be cast as a sparse signal
recovery problem and existing compressed-sensing methods can be
employed. Simulation results show that our proposed method, with
only a small amount of training overhead, can provide reliable
channel estimation and help attain a decent beamforming gain.

\section{System Model and Problem Formulation}
We consider an IRS-assisted mmWave downlink system, where an IRS
is deployed to assist the data transmission from the BS to a
single-antenna user. Suppose the IRS is a planar array with $M$
reflecting elements. The BS is equipped with $N$ antennas. Let
$\boldsymbol{G} \in \mathbb C^{M \times N}$ denote the channel
from the BS to the IRS, and $\boldsymbol{h}_r \in \mathbb C^{M}$
denote the channel from the IRS to the user. To better illustrate
our idea, we neglect the direct link from the BS to the user.
Nevertheless, the extension to the scenario with direct link from
the BS to the user is straightforward. Each reflecting element of
the IRS can reflect the incident signal with a reconfigurable
phase shift and amplitude via a smart controller
\cite{WuZhang19b}. Denote
\begin{align}
\boldsymbol{\Theta} \triangleq\text{diag}(\beta_{1}
e^{j\theta_{1}},\ldots,\beta_{M} e^{j\theta_{M}})
\end{align}
as the phase-shift matrix of the IRS, where $\theta_{m} \in
[0,2\pi]$ and $\beta_{m} \in [0,1]$ denote the phase shift and
amplitude reflection coefficient associated with the $m$th passive
element of the IRS, respectively. For simplicity, we assume
$\beta_m =1,\forall m$ in the sequel of this paper.

Let $\boldsymbol{w}\in \mathbb{C}^{N}$ denote the beamforming
vector adopted by the BS. The signal received by the user at the
$t$th time instant is given by
\begin{align}
y(t) &=  \boldsymbol{h}_{r}^H
\boldsymbol{\Theta}(t) \boldsymbol{G} \boldsymbol{w}(t) s(t) + \epsilon (t)\nonumber \\
& \stackrel{(a)}= \boldsymbol{v}^H(t) \text {diag}( \boldsymbol{h}_r^H) \boldsymbol{G}
\boldsymbol{w}(t) s(t) + \epsilon(t) \nonumber \\
& \stackrel{(b)} = \boldsymbol{v}^H(t) \boldsymbol{H}
\boldsymbol{w}(t) s(t) + \epsilon(t) \label{re-sig-noD}
\end{align}
where $s(t)$ is the transmitted symbol, $\epsilon(t)$ denotes the
additive white Gaussian noise with zero mean and variance
$\sigma^2$, in $(a)$, we define $\boldsymbol{v}\triangleq
[e^{j\theta_1}\phantom{0}\ldots\phantom{0} e^{j\theta_M}]^H \in
\mathbb C^{M}$, and in $(b)$, we define
$\boldsymbol{H}\triangleq\text {diag}(
\boldsymbol{h}_r^H)\boldsymbol{G}$. Here $\boldsymbol{H}$ is
referred to as the cascade channel. An important observation based
on (\ref{re-sig-noD}) is that, in the beamforming stage, we only
need the knowledge of the cascade channel $\boldsymbol{H}$ for
joint active and passive beamforming, i.e. optimizing
$\boldsymbol{w}$ and $\boldsymbol{v}$ to maximize the received
signal power at the receiver. Therefore, in the channel estimation
stage, our objective is to estimate the cascade channel
$\boldsymbol{H}$ from the received measurements
$\{y(t)\}_{t=1}^T$. Note that to facilitate channel estimation,
different precoding vectors $\{\boldsymbol{w}(t)\}$ are employed
at different time instants, while the phase shift vector
$\boldsymbol{v}$ can either be time-varying or remain
time-invariant over different time instants. Without loss of
generality, we use $\boldsymbol{v}(t)$ to represent the phase
shift vector used at the $t$th time instant. We also would like to
clarify that the channel estimation algorithm is implemented at
the receiver (i.e. user), no operation or algorithm needs to be
executed at the IRS.

The cascade channel matrix $\boldsymbol{H}$ has a dimension of
$M\times N$. Both $N$ and $M$ could be large for mmWave systems,
which makes channel estimation a challenging problem. Hopefully,
real-world channel measurements \cite{RappaportSun13,AkdenizLiu14}
have shown that mmWave channels exhibit sparse scattering
characteristics, which can be utilized to substantially reduce the
training overhead.

\section{Channel Model}
Following \cite{AlkhateebAyach14}, a narrowband geometric channel
model is used to characterize the BS-IRS channel $\boldsymbol{G}$
and the IRS-user channel $\boldsymbol{h}_r$. Specifically, the
BS-IRS channel can be modeled as
\begin{align}
\boldsymbol{G} =
\sqrt{\frac{NM}{\rho}}\sum_{l=1}^L\varrho_l\boldsymbol{a}_r(\vartheta_l,\gamma_l)
\boldsymbol{a}_{t}^H(\phi_l)
\end{align}
where $\rho$ denotes the average path-loss between the BS and IRS,
$L$ is the number of paths, $\varrho_l$ denotes the complex gain
associated with the $l$th path, $\vartheta_{l}$ ($\gamma_{l}$)
denotes the azimuth (elevation) angle of arrival (AoA), $\phi_l$
is the angle of departure (AoD), $\boldsymbol{a}_r$ and
$\boldsymbol{a}_{t}$ represent the receive and transmit array
response vectors, respectively. Suppose the IRS is an $M_x \times
M_y$ uniform planar array (UPA). We have \cite{DingRao18}
\begin{align}
\boldsymbol{a}_r(\vartheta_l,\gamma_l)= \boldsymbol{a}_{x}(u)
\otimes \boldsymbol{a}_{y}(v)
\end{align}
where $\otimes$ stands for the Kronecker product, $u\triangleq 2
\pi d \cos( \gamma_l) / \lambda$, $v\triangleq 2 \pi d
\sin(\gamma_l) \cos( \vartheta_l) / \lambda$, $d$ denotes the
antenna spacing, $\lambda$ is the signal wavelength, and
\begin{align}
\boldsymbol{a}_{x}(u) \triangleq&
\frac{1}{\sqrt{M_x}} [1\phantom{0}e^{ju}\phantom{0}\ldots\phantom{0} e^{j(M_x-1)u}]^T
\nonumber\\
\boldsymbol{a}_{y}(v) \triangleq&
 \frac{1}{\sqrt{M_y}}  [1\phantom{0}e^{jv}\phantom{0}\ldots\phantom{0} e^{j(M_y-1)v}]^T
\end{align}
Due to the sparse scattering nature of mmWave channels, the number
of path $L$ is small relative to the dimensions of
$\boldsymbol{G}$. Hence we can express $\boldsymbol{G}$ as
\begin{align}
\boldsymbol{G} = (\boldsymbol{F}_{x} \otimes \boldsymbol{F}_{y})
\boldsymbol{\Sigma} \boldsymbol{F}_L^H \triangleq
\boldsymbol{F}_P\boldsymbol{\Sigma} \boldsymbol{F}_L^H
\label{ch-G}
\end{align}
where $\boldsymbol{F}_L \in \mathbb C^{N \times N_G}$ is an
overcomplete matrix ($N_G\geq N$) and each of its columns has a
form of $\boldsymbol{a}_{t}(\phi_l)$, with $\phi_l$ chosen from a
pre-discretized grid, $\boldsymbol{F}_x\in\mathbb{C}^{M_x \times
M_{G,x}}$ ($\boldsymbol{F}_y\in \mathbb{C}^{M_y \times M_{G,y}}$)
is similarly defined with each of its columns having a form of
$\boldsymbol{a}_{x}(u)$ ($\boldsymbol{a}_{y}(v)$), and $u$ ($v$)
chosen from a pre-discretized grid,
$\boldsymbol{\Sigma}\in\mathbb{C}^{M_G\times N_G}$ is a sparse
matrix with $L$ non-zero entries corresponding to the channel path
gains $\{\varrho_l\}$, in which $M_G=M_{G,x}\times M_{G,y}$. Here
for simplicity, we assume that the true AoA and AoD parameters lie
on the discretized grid. In the presence of grid mismatch, the
number of nonzero entries will become larger due to the power
leakage caused by the grid mismatch
\cite{ChiScharf11}.

The IRS-user channel can be modeled as
\begin{align}
\boldsymbol{h}_r =\sqrt{\frac{M}{\varepsilon}} \sum_{l=1}^{L'} \alpha_l
\boldsymbol{a}_r(\vartheta_l,\gamma_l)
\end{align}
where $\varepsilon$ denotes the average path-loss between the IRS and the user,
$\alpha_l$ denotes the complex gain associated with the
$l$th path, and $\vartheta_{l}$ ($\gamma_{l}$) denotes the azimuth
(elevation) angle of departure. Due to limited scattering
characteristics, the IRS-user channel can be written as
\begin{align}
\boldsymbol{h}_r &= \boldsymbol{F}_P \boldsymbol{\alpha}
\label{ch-r}
\end{align}
where $\boldsymbol{\alpha}\in\mathbb{C}^{M_G}$ is a sparse vector
with $L'$ nonzero entries.

\section{Proposed Method}
\subsection{Channel Estimation}
We now discuss how to develop a compressed sensing-based method to
estimate the cascade channel $\boldsymbol{H}$. Let $\bullet$
denote the ``transposed Khatri-Rao product'', we can express the
cascade channel as
\begin{align}
\boldsymbol{H} & =\text {diag}( \boldsymbol{h}_r^H) \boldsymbol{G}
\stackrel{(a)}= \boldsymbol{h}_r^{\ast} \bullet \boldsymbol{G}  \nonumber \\
& \stackrel{(b)}=  (\boldsymbol{F}_P^{\ast} \boldsymbol{\alpha}^{\ast})
\bullet ( \boldsymbol{F}_P  \boldsymbol{\Sigma} \boldsymbol{F}_L^H ) \nonumber \\
& \stackrel{(c)}=  (\boldsymbol{F}_P^{\ast} \bullet \boldsymbol{F_P})
( \boldsymbol {\alpha}^{\ast} \otimes (\boldsymbol{\Sigma} \boldsymbol{F}_L^H)) \nonumber \\
& \stackrel{(d)}= (\boldsymbol{F}_P^{\ast} \bullet \boldsymbol{F}_P)
( \boldsymbol {\alpha}^{\ast} \otimes \boldsymbol {\Sigma}  ) ( 1 \otimes \boldsymbol{F}_L^H ) \nonumber \\
& \stackrel{(e)}=\boldsymbol{D}( \boldsymbol {\alpha}^{\ast}
\otimes \boldsymbol {\Sigma}  )  \boldsymbol{F}_L^H \label{sp-H}
\end{align}
where in $(a)$, $(\cdot)^{\ast}$ denotes the complex conjugate,
$(b)$ comes from \eqref{ch-G} and \eqref{ch-r}, $(c)$ follows from
the property of Khatri-Rao product (see (1.10.27) in
\cite{Zhang17}), $(d)$ is obtained by resorting to the property of
Kronecker product (see (1.10.4) in \cite{Zhang17}), and we define
$\boldsymbol{D} \triangleq \boldsymbol{F}_P^{\ast} \bullet
\boldsymbol{F}_P$ in $(e)$. Since both $\boldsymbol{\alpha}$ and
$\boldsymbol{\Sigma}$ are sparse, their Kronecker product is also
sparse. We see that after a series of transformation, a sparse
representation of the cascade channel $\boldsymbol{H}$ is
obtained. This sparse formulation can be further simplified by
noticing that the matrix $\boldsymbol{D}$ contains a considerable
amount of redundant columns due to the transposed Khatri-Rao
product operation. Specifically, we have the following result
regarding the redundancy of $\boldsymbol{D}$.

\newtheorem{proposition}{Proposition}
\begin{proposition}
The matrix $\boldsymbol{D}\in \mathbb C^{M \times M_G^2}$ only
contains $M_G$ distinct columns which are exactly the first $M_G$
columns of $\boldsymbol{D}$, i.e.
\begin{align}
 \boldsymbol{D}_u =\boldsymbol{D}(:, 1:M_G)
\end{align}
where $\boldsymbol{D}_u$ denotes a matrix constructed by the $M_G$
distinct columns of $\boldsymbol{D}$. \label{proposition1}
\end{proposition}

The proof is easily verified and thus omitted. Based on this
result, the cascade channel $\boldsymbol{H}$ can be further
expressed as
\begin{align}
\boldsymbol{H} =  \boldsymbol{D}( \boldsymbol {\alpha}^{\ast}
\otimes \boldsymbol {\Sigma}  )  \boldsymbol{F}_L^H  =
\boldsymbol{D}_u \boldsymbol{\Lambda}  \boldsymbol{F}_L^H
\label{sp-Hu}
\end{align}
where $\boldsymbol{\Lambda} \in \mathbb C^{ M_G \times N_G}$ is a
merged version of $(\boldsymbol {\alpha}^{\ast} \otimes
\boldsymbol {\Sigma}) \triangleq \boldsymbol{J}$, with each of its
rows being a superposition of a subset of rows in
$\boldsymbol{J}$, i.e. $\boldsymbol{\Lambda}(i,:) = \sum_{n \in
\mathcal{S}_i} \boldsymbol{J} (n,:)$, where
$\boldsymbol{\Lambda}(i,:)$ denotes the $i$th row of
$\boldsymbol{\Lambda}$, $\mathcal{S}_i$ denotes the set of indices
associated with those columns in $\boldsymbol{D}$ that are
identical to the $i$th column of $\boldsymbol{D}$. Clearly, there
are at most $L\times L'$ nonzero entries in
$\boldsymbol{\Lambda}$.

Assuming the pilot signal $s(t)=1,\forall t$, the received signal
$y(t)$ in (\ref{re-sig-noD}) can be written as
\begin{align}
 y(t) = &  \boldsymbol{v}^H(t) \boldsymbol{H} \boldsymbol{w}(t) s(t) + \epsilon (t)\nonumber \\
 \stackrel{(a)} = & \left(  \boldsymbol{w}^T(t)\otimes  \boldsymbol{v}^H(t) \right)
 \text{vec} (\boldsymbol{H} )+\epsilon(t) \nonumber \\
\stackrel{(b)} =&   \left(  \boldsymbol{w}^T(t)\otimes
\boldsymbol{v}^H(t) \right)  \left(
 \boldsymbol{F}_L^{\ast} \otimes  \boldsymbol{D}_u\right)
 \text{vec}(\boldsymbol{\Lambda})+\epsilon(t)\nonumber \\
\stackrel{(c)} =& \left(  \boldsymbol{w}^T(t)\otimes
\boldsymbol{v}^H(t) \right)  \boldsymbol{\tilde{F}} \boldsymbol{x}
+\epsilon(t) \label{sp-y}
 \end{align}
where $(a)$ and $(b)$ follow from the property of Kronecker
product, and in $(c)$ we define $\boldsymbol{\tilde{F}}\triangleq
\boldsymbol{F}_L^{\ast} \otimes \boldsymbol{D}_u$ and
$\boldsymbol{x} \triangleq \text{ vec}( \boldsymbol{\Lambda})$.
Stacking the measurements collected at different time instants
$\boldsymbol{y} \triangleq [ y(1) \phantom{0} \ldots \phantom{0}
y(T)]^T$, we arrive at
\begin{align}
\boldsymbol{y} = \boldsymbol{\Phi} \boldsymbol{x} +
\boldsymbol{\epsilon} \label{sp-vecy}
\end{align}
where
$\boldsymbol{\Phi}\triangleq\boldsymbol{W}_v\boldsymbol{\tilde{F}}$
and
\begin{align}
\boldsymbol{W}_v\triangleq\left[
\begin{matrix}
 \boldsymbol{w}^T(1)\otimes  \boldsymbol{v}^H(1) \\
 \vdots \\
  \boldsymbol{w}^T(T)\otimes  \boldsymbol{v}^H(T)
\end{matrix}
\right]
\end{align}

So far we have converted the channel estimation problem into a
sparse signal recovery problem, and many classical compressed
sensing algorithms such as the orthogonal matching pursuit (OMP)
\cite{TroppGilbert07} can be
employed to estimate the sparse signal $\boldsymbol{x}$. After
$\boldsymbol{x}$ is recovered, the cascade channel
$\boldsymbol{H}$ can be accordingly obtained via (\ref{sp-Hu}).

In the following, we analyze the sample complexity of our proposed
compressed sensing-based method. According to the compressed
sensing theory, for an underdetermined system of linear equations
$\boldsymbol{y}=\boldsymbol{Ax}$, the number of measurements
required for successful recovery of $\boldsymbol{x}$ is at the
order of $O(k\log n)$, where $n$ is the dimension of
$\boldsymbol{x}$, and $k$ denotes the number of nonzero elements
in $\boldsymbol{x}$. For the sparse signal recovery problem
(\ref{sp-vecy}), we have $n=M_G N_G$ and $k\leq L L'$. Therefore
our proposed method has a sample complexity of $\mathcal{O}(L L'
\log(M_G N_G))$. Due to the sparse scattering nature of mmWave
channels, $LL'$ is much smaller than $MN$. Therefore a substantial
training overhead reduction can be achieved.

\begin{figure*}[!t]
\centering
\subfigure[NMSE.]{\includegraphics[width=5.4cm]{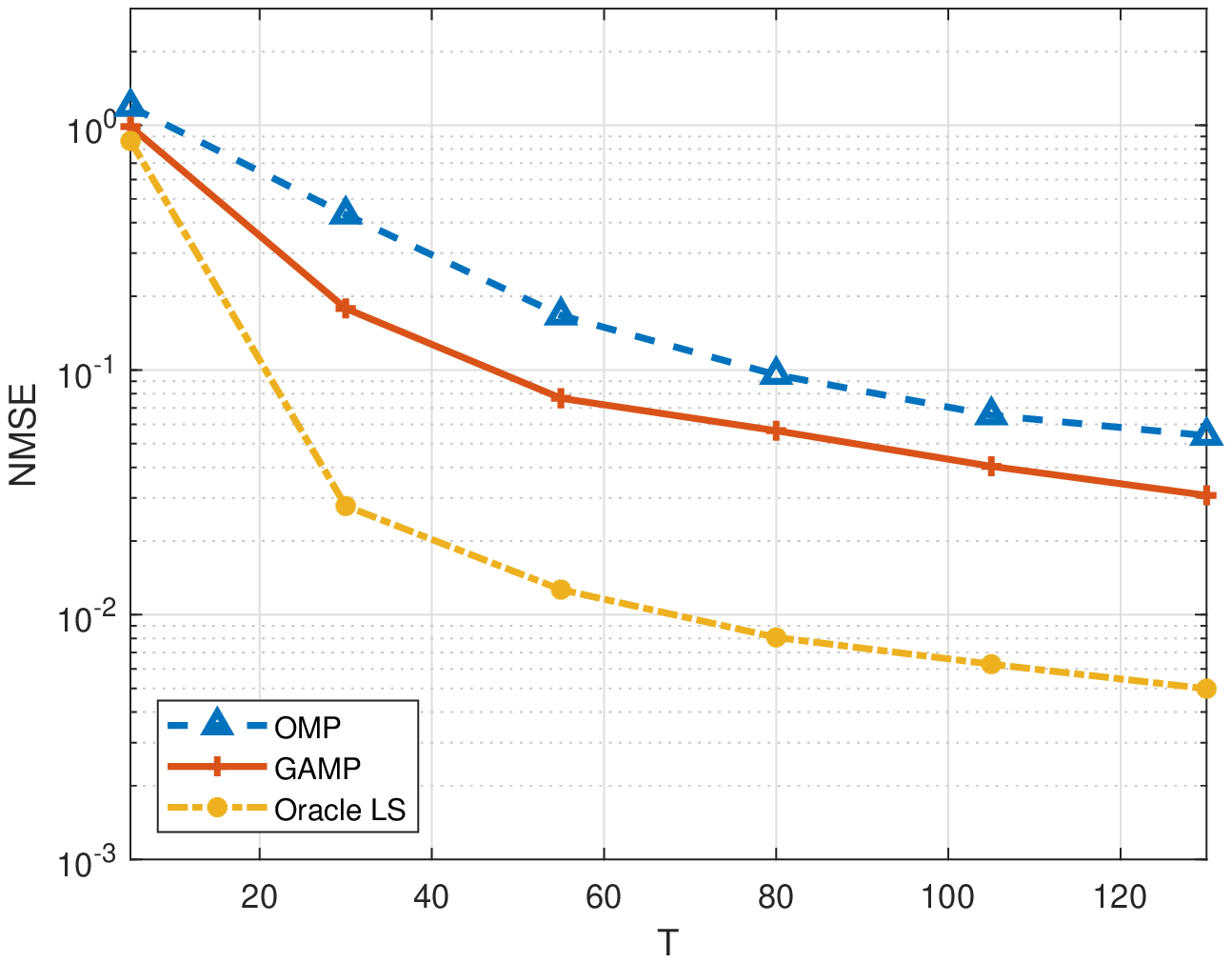}} \hfil
\subfigure[ARSPR.]{\includegraphics[width=5.4cm]{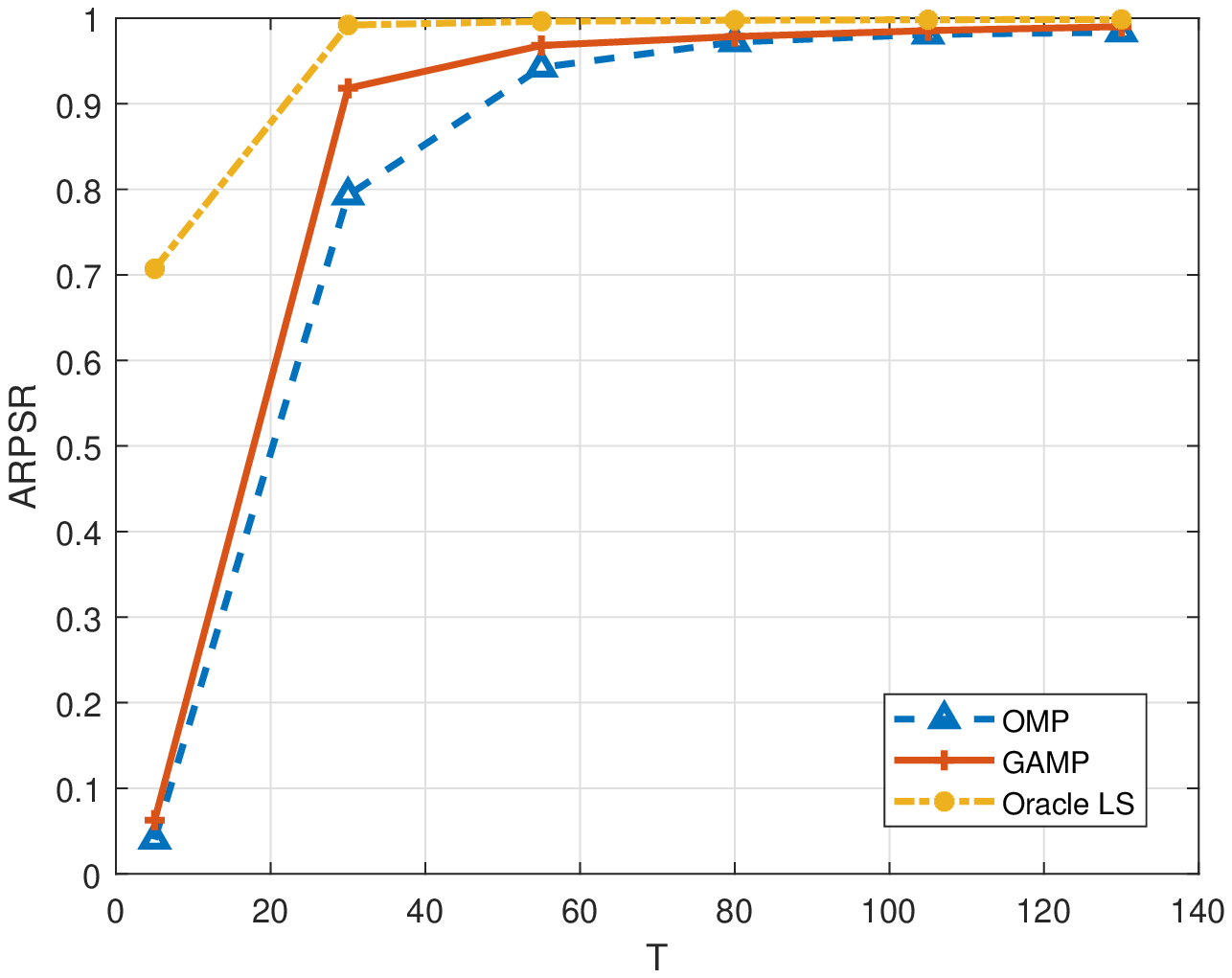}}
\caption{NMSEs and ARSPRs of respective algorithms vs. $T$}
\label{fig1}
\end{figure*}

\begin{figure*}[!t]
\centering
\subfigure[NMSE.]{\includegraphics[width=5.4cm]{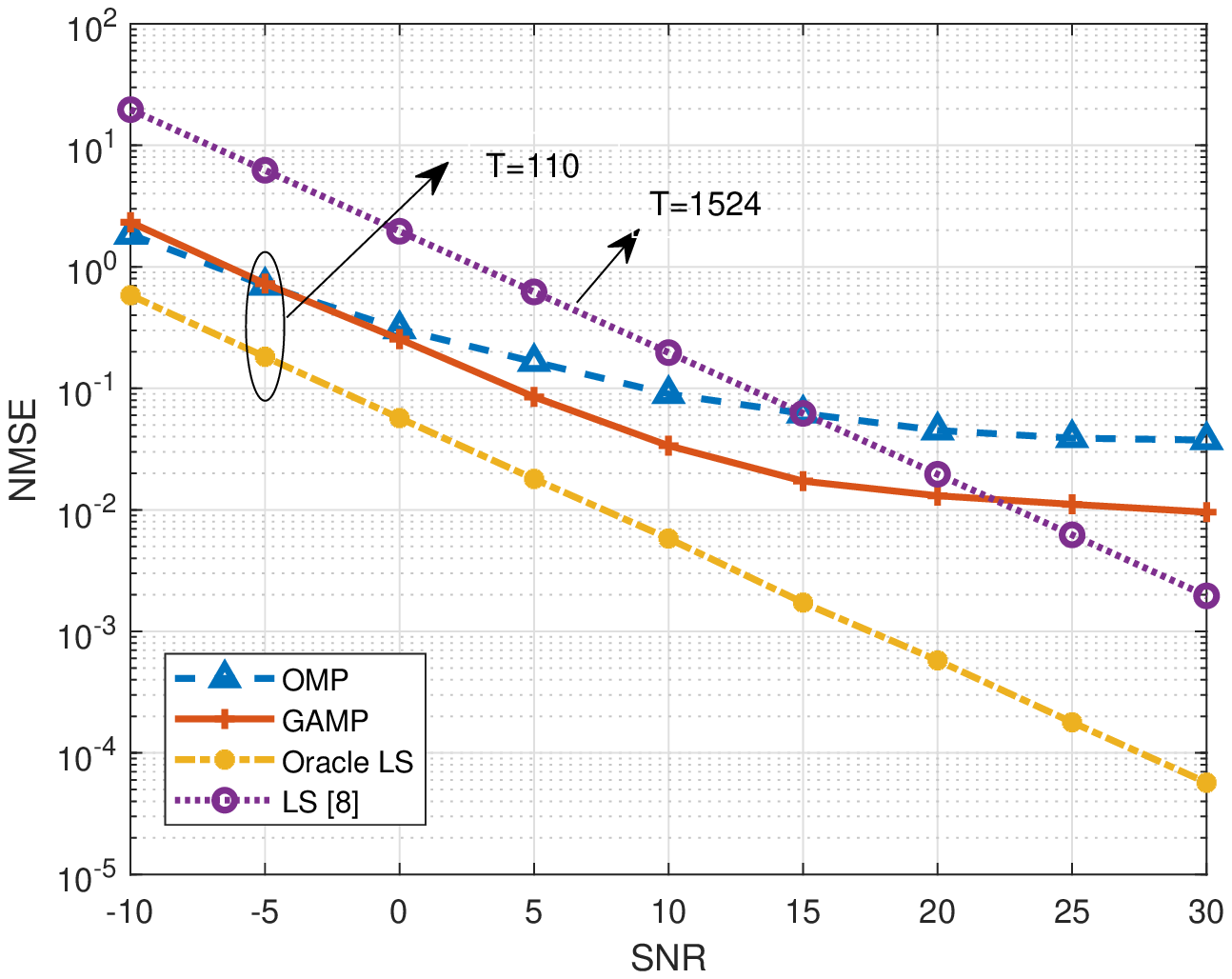}} \hfil
\subfigure[ARSPR. ]{\includegraphics[width=5.4cm]{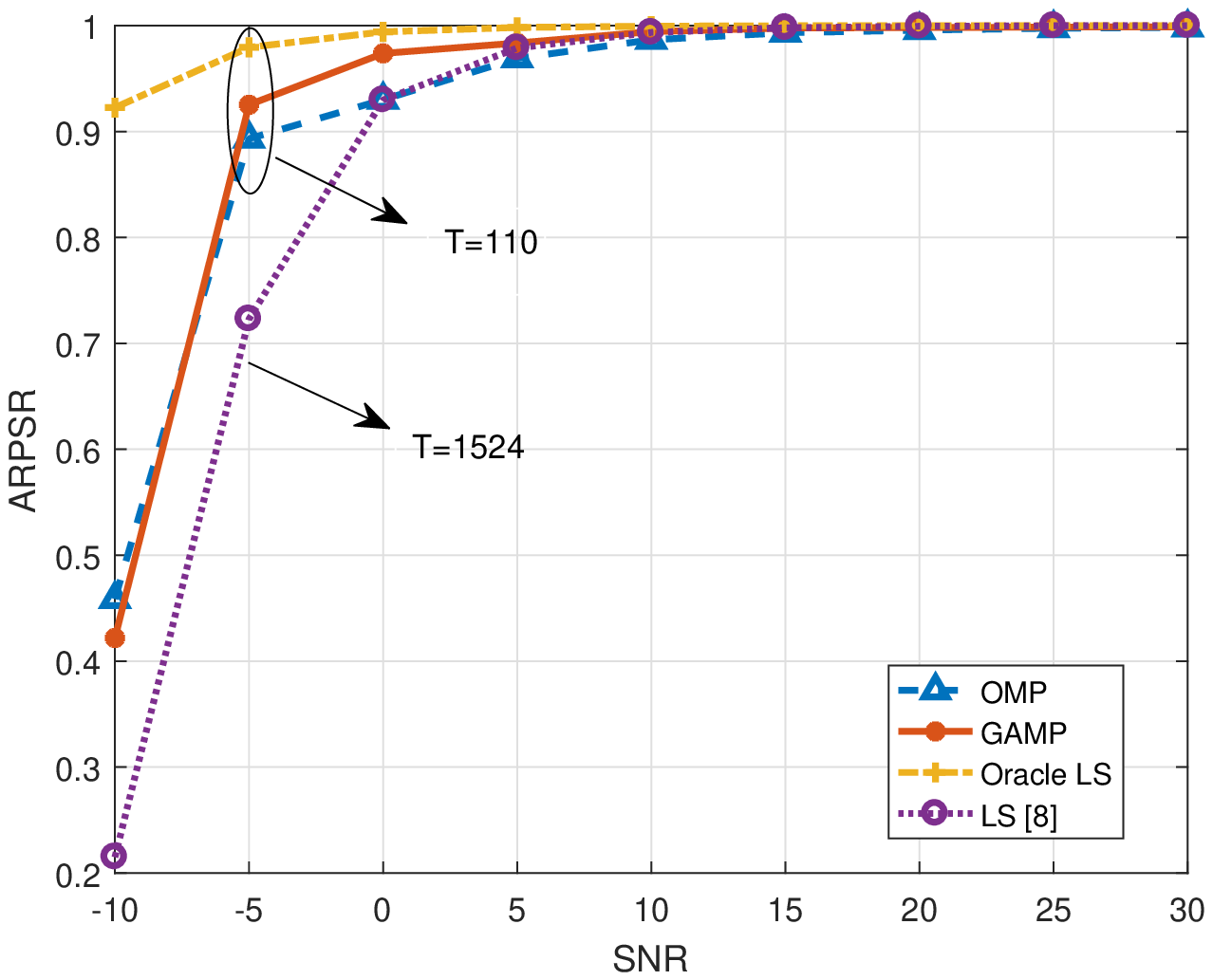}}
\caption{NMSEs and ARSPRs of respective algorithms vs. SNR}
\label{fig2}
\end{figure*}

\subsection{Extension To Multi-Antenna Receiver}
We briefly discuss the extension of the proposed method to the
multi-antenna receiver scenario. Suppose the user is equipped with
$N_r$ antennas. Let $\boldsymbol{R}\in\mathbb{C}^{N_r\times M}$
denote the channel from the IRS to the user. Due to the sparse
scattering nature of mmWave channels, we can write $\boldsymbol{R}
= \boldsymbol{F}_r \boldsymbol{\Gamma} \boldsymbol{F}_P^H$ where
$\boldsymbol{F}_r \in \mathbb C^{N_r \times N_{G_r}}$ is an
overcomplete matrix and each of its columns has a form of
$\boldsymbol{a}_r( \phi_i)$, with $\phi_i$ chosen from a
pre-discretized grid, and $\boldsymbol{\Gamma} \in \mathbb
C^{N_{G_r} \times M_G}$ is a sparse matrix with $L'$ nonzero
elements. For the MIMO scenario, the signal received by the user
at the $t$th time instant is given by
\begin{align}
y(t)=\boldsymbol{f}^H(t)\boldsymbol{\bar
H}\boldsymbol{w}(t)s(t)+\epsilon(t)
\end{align}
where $\boldsymbol{f}(t)$ and $\boldsymbol{w}(t)$ denote the
combining vector at the receiver and the precoding vector at the
transmitter respectively, and $\boldsymbol{\bar H}$ is defined as
\begin{align}
\boldsymbol{\bar H} & \triangleq \boldsymbol{R}
\boldsymbol{\Theta}\boldsymbol{G} = \boldsymbol{F}_r
\boldsymbol{\Gamma} \boldsymbol{F}_P^H \boldsymbol{\Theta}
\boldsymbol{F}_{P} \boldsymbol{\Sigma} \boldsymbol{F}_L^H \nonumber \\
& \triangleq \boldsymbol{F}_r \boldsymbol{\Gamma} \boldsymbol{\Xi}
\boldsymbol{\Sigma} \boldsymbol{F}_L^H
\end{align}
in which $\boldsymbol{\Xi} \triangleq  \boldsymbol{F}_P^H
\boldsymbol{\Theta}\boldsymbol{F}_{P}$. Furthermore, we have
\begin{align}
{\rm vec} (\boldsymbol{\bar H}) = &{\rm vec} ( \boldsymbol{F}_r
\boldsymbol{\Gamma} \boldsymbol{\Xi} \boldsymbol{\Sigma}
\boldsymbol{F}_L^H ) =  (\boldsymbol{F}_L^{\ast} \otimes
\boldsymbol{F}_r) {\rm vec}
(  \boldsymbol{\Gamma} \boldsymbol{\Xi} \boldsymbol{\Sigma}) \nonumber \\
= & (\boldsymbol{F}_L^{\ast} \otimes \boldsymbol{F}_r)
( \boldsymbol{ \Sigma}^T \otimes  \boldsymbol{ \Gamma}) {\rm vec}( \boldsymbol{\Xi}) \nonumber \\
= &  (\boldsymbol{F}_L^{\ast} \otimes \boldsymbol{F}_r) (
\boldsymbol{ \Sigma}^T \otimes  \boldsymbol{ \Gamma}) (
\boldsymbol{F}_P^T\odot\boldsymbol{F}_P^H) \boldsymbol{v}
^{\ast}\label{vecH-mimo}
\end{align}
where $\odot$ denotes the Khatri-Rao product. Similarly, the
matrix $\boldsymbol{\bar D}  \triangleq
\boldsymbol{F}_P^T\odot\boldsymbol{F}_P^H \in \mathbb C^{M_G^2
\times M}$  contains only $M_G$ distinct rows which are exactly
the first $M_G$ rows of $\boldsymbol{\bar D}$. Thus we can rewrite
\eqref{vecH-mimo} as
\begin{align}
{\rm vec} (\boldsymbol{\bar H})  = (\boldsymbol{F}_L^{\ast}
\otimes \boldsymbol{F}_r) \boldsymbol{ \bar \Lambda}
\boldsymbol{\bar D}_u \boldsymbol{v}^{\ast}
\end{align}
where $ \boldsymbol{\bar D}_u \triangleq \boldsymbol{\bar D}
(1:M_G,:)$, $\boldsymbol{\bar \Lambda} $ is a merged version of
$\boldsymbol{\bar J} \triangleq \boldsymbol{ \Sigma}^T \otimes
\boldsymbol{ \Gamma}$, i.e. $\boldsymbol{\bar \Lambda}(:,i) =
\sum_{n \in \mathcal{Q}_i} \boldsymbol{\bar J}(:,n)$, where
$\mathcal{Q}_i$ is the set of indices associated with those rows
in $\boldsymbol{\bar D}$ that are identical to the $i$th row of
$\boldsymbol{\bar D}$. Hence, we can move on to write
\begin{align}
\rm vec(\boldsymbol{\bar H} ) =   & (\boldsymbol{F}_L^{\ast}
\otimes \boldsymbol{F}_r) \boldsymbol{ \bar \Lambda}
\boldsymbol{\bar D}_u \boldsymbol{v}^{\ast}  \nonumber \\
=& \left( ( \boldsymbol{\bar D}_u \boldsymbol{v}^{\ast} )^T
\otimes (\boldsymbol{F}_L^{\ast} \otimes
\boldsymbol{F}_r) \right) {\rm vec}( \boldsymbol{\bar \Lambda}) \nonumber \\
 \triangleq & \boldsymbol{K} \boldsymbol{\bar x}
\end{align}
where $\boldsymbol{K} \triangleq( \boldsymbol{\bar D}_u
\boldsymbol{v}^{\ast} )^T \otimes (\boldsymbol{F}_L^{\ast} \otimes
\boldsymbol{F}_r)  $ and $\boldsymbol{\bar x} \triangleq {\rm
vec}( \boldsymbol{\bar \Lambda}) $ is a sparse vector to be
estimated. Let $s(t)=1$, and define $\boldsymbol{y} =
[y(1)\phantom{0}y(2)\phantom{0} \ldots\phantom{0} y(T)]^T$, we
have
\begin{align}
\boldsymbol{y} = \boldsymbol{W}_f \boldsymbol{K} \boldsymbol{\bar
x} +\boldsymbol{\epsilon} \label{eqn1}
\end{align}
where $\boldsymbol{W}_f \in \mathbb C^{T \times N M}$,
$\boldsymbol{W}_f(t,:) = \boldsymbol{w}^T(t) \otimes
\boldsymbol{f}^H(t)$, and $\boldsymbol{W}_f(t:)$ is the $t$th row
of $\boldsymbol{W}_f$. We see that estimation of the channel
vector $\boldsymbol{\bar x}$ is converted to a conventional sparse
signal recovery problem. Note that although we cannot obtain
$\boldsymbol{\Gamma}$ (i.e. $\boldsymbol{R}$) and
$\boldsymbol{\Sigma}$ (i.e. $\boldsymbol{G}$) from
$\boldsymbol{\bar x}$, the knowledge of $\boldsymbol{\bar x}$
itself is enough for joint beamforming for the MIMO scenario as
the joint beamforming problem can be converted to an optimization
problem which maximizes $\| \boldsymbol{\bar H}\|_F^2=\| {\rm vec}
(\boldsymbol{\bar H})\|_2^2$ with respect to $\boldsymbol{v}$
\cite{NingChen20,ZhangZhang19}.

\section{Simulation Results} \label{sec:experiments}
In this section, we present simulation results to evaluate the
performance of our proposed channel estimation method. Two
different compressed sensing algorithms, namely, the OMP
\cite{TroppGilbert07} and GAMP \cite{VilaSchniter13} are employed
to solve (\ref{sp-vecy}). To provide a benchmark for our proposed
method, we compare with the oracle least-squares (Oracle-LS)
estimator which assumes the knowledge of the support of the sparse
signal. Clearly, the oracle LS estimator provides the best
achievable performance for any compressed sensing-based method.
Also, we compare with the conventional LS estimator proposed in
\cite{MishraJohansson19}, which sets $T\geq NM$ and formulates
channel estimation as an over-determined system of equations:
$\boldsymbol{y} = \boldsymbol{W}_v {\rm vec}(\boldsymbol{H}) +
\boldsymbol{\epsilon}$ (cf. (\ref{sp-y})). We assume that the BS
employs a uniform linear array (ULA) with $N=16$ antennas and the
IRS is a UPA consisting of $M=8\times 8$ passive reflecting
elements. In our simulations, we set $N_G=64$, $M_{G,x}=32$ and
$M_{G,y}=32$. Also, we assume a Rician channel comprising a LOS
path and a number of NLOS paths
\cite{HurKim13,LiFang19,Muhi-EldeenIvrissimtzis10}. The Rician
factor is set to $13.2$dB according to
\cite{Muhi-EldeenIvrissimtzis10}. The number of paths for mmWave
channels $\boldsymbol{G}$ and $\boldsymbol{h}_r$ are respectively
set to $L=3$ and $L'=3$, where the AoA and AoD parameters are
uniformly generated from $[-\pi/2,\pi/2]$ and not necessarily lie
on the discretized grid.

The performance is evaluated via two metrics, i.e. normalized mean
squared error (NMSE) and average receive signal power ratio
(ARSPR). The NMSE is defined as $\mathbb E [ \|(\boldsymbol{ \hat
H} - \boldsymbol{ H} ) \| _F^2/ \| \boldsymbol{H} \|_F^2]$. The
ARSPR is defined as the ratio of the actual receive signal power
to the ideal receive signal power, i.e. $\mathbb E[\|\boldsymbol{
v}^H \boldsymbol{H} \|_F^2 / \| (\boldsymbol{v}^{\star})^H
\boldsymbol{H}\|_F^2]$, where $\boldsymbol{v}$ and
$\boldsymbol{v}^{\star}$ are respectively obtained via solving the
joint beamforming problem \cite{YuXu19} based on the estimated
cascade channel $\boldsymbol{ \hat {H}}$ and the real channel
$\boldsymbol{H}$. In Fig. \ref{fig1}, we plot the NMSEs and ARSPRs
of respective algorithms as a function of $T$, where the
signal-to-noise ratio (SNR) is set as $10$dB. From Fig.
\ref{fig1}, we see that GAMP only needs about $100$ measurements
to attain an NMSE as low as $0.04$, thus achieving a substantial
overhead reduction. Fig. \ref{fig2} depicts the NMSEs and ARSPRs
versus the SNR, where we set $T=110$ for GAMP, OMP and the oracle
LS estimator, and $T=1524>NM$ for the conventional LS estimator
\cite{MishraJohansson19}. Our results show that our proposed
method can achieve performance (in terms of ARPSR) close to that
of the oracle LS estimator in scenarios of practical interest,
e.g. $T>40$ and $\text{SNR}>-5\text{dB}$. Meanwhile, it can be
observed that the conventional LS estimator
\cite{MishraJohansson19} requires much more measurements to
achieve a performance similar to the proposed method. Also, the
conventional LS estimator has a computational complexity of
$\mathcal{O}(T(NM)^2)$ with $T \geq NM$, while the complexity of
OMP and GAMP is of $\mathcal{O}(TNM)$ with $T < NM$. Hence our
proposed method is more computationally efficient than the
conventional LS estimator \cite{MishraJohansson19}.

\section{Conclusions}
We studied the problem of channel estimation and joint beamforming
design for IRS-assisted mmWave systems. We proposed a compressed
sensing-based channel estimation method by exploiting the inherent
sparse struccture of the cascade channel. Simulation results
showed that our proposed method can provide an accurate channel
estimation and achieve a substantial training overhead reduction.

%\bibliography{newbib}
%\bibliographystyle{IEEEtran}

\end{document}